\documentclass{svjour3}                     
\smartqed  
\usepackage{graphicx}
\usepackage{amsmath}
\usepackage{mathptmx}      
\newcommand{\rb}{{\bf r}}
\newcommand{\rp}{{\bf r'}}
\newcommand{\co}{{\bf c}}
\newcommand{\dr}{\,d{\bf r}}

\begin{document}

\title{Brief review related to the foundations of time-dependent density functional theory}


\titlerunning{Foundations of TDDFT}        

\author{Thomas A. Niehaus  \and
        Norman H. March
}
\dedication{dedicated to Professor S\'{a}ndor Suhai on the occasion of his 65th birthday}

\institute{T.A. Niehaus \at
              Bremen Center for Computational Materials Science, 
          D-28359 Bremen, Germany \\
              \email{t.niehaus@bccms.uni-bremen.de}         
           \and
           N.H. March \at
              Dept. of Physics, University of Antwerp, Antwerp, Belgium\\
Oxford University, Oxford, England
}

\date{Received: date / Accepted: date}

\maketitle

\begin{abstract}
The electron density $n(\rb,t)$, which is the central tool of time-dependent density functional theory, is presently considered to be derivable from a one-body time-dependent potential $V(\rb,t)$, via one-electron wave functions satisfying a time- dependent Schr\"{o}dinger equation. This is here related via a generalized equation of motion to a Dirac density matrix now involving $t$. Linear response theory is then surveyed, with a special emphasis on the question of causality with respect to the density dependence of the potential. Extraction of $V(\rb,t)$ for solvable models is also proposed. 
\end{abstract}

\section{Background}
\label{intro}
Early work on time-dependent density functional theory can be traced
back at least to 1972 \cite{Mar72,Mar73,Gro90} followed by various
studies \cite{And77,And77_2,Zan80,Zan81} in the 1970s / early 1980s,
and culminating in the important proposal of Runge and Gross (RG)
\cite{Run84}. In essence, the RG argument generalizes the
Hohenberg-Kohn theorem \cite{Hoh64} to time-dependent external
potentials. Though parts of this important study were questioned in \cite{Sch07}, the RG work is widely accepted as the basis for the assertion that, for a specified initial state, there is a unique correspondence between the time dependent density $n(\rb,t)$ and the external potential $V_{\text{ext}}(\rb,t)$. This points the way to construct a time-dependent extension $V(\rb,t)$ of a static Slater-Kohn-Sham (SKS) like potential $V(\rb)$ \cite{Sla51,Koh65} which will then generate a Slater determinant of non-interacting electron wave functions, $\phi_i(\rb,t)$ say, satisfying the time-dependent Schr\"{o}dinger equation
\begin{equation}
\label{tdse}
\left( -\frac{\hbar}{2m} \nabla^2 + V(\rb,t) \right) \phi_i(\rb,t) = i\hbar \frac{\partial}{\partial t} \phi_i(\rb,t),
\end{equation} 
with a specified determinant at time $t=0$. Then the electron density $n(\rb,t)$  introduced above is constructed, formally exactly for N electrons, as
\begin{equation}
\label{tdden}
n(\rb,t) = \sum_i^N \phi_i^*(\rb,t)\phi_i(\rb,t).
\end{equation}
Of course, formal exactitude requires precise knowledge of the
one-body potential $V(\rb,t)$ in eqn(\ref{tdse}). At the time of
writing, such knowledge is limited for the key exchange and
correlation contributions entering $V$ (see also eqn(\ref{fxc})
below). These can be formally constructed \cite{Qia98,Qia00,Qia01}, and
perturbative approaches that converge on their exact form are also
known
\cite{PhysRevLett.51.1888,PhysRevLett.76.3610,PhysRevA.50.196}. In
practical terms, however, accurate
exchange-correlation functionals beyond the local density
approximation in space and time are still elusive and remain a topic of general interest.
\section{The challenge of Schirmer and Dreuw \cite{Sch07} to the RG arguments, and some responses}
This is the point to return to the work of \cite{Sch07}. This study
contains serious criticism levelled against the very foundations of
TDDFT (articles \cite{Xu85,Dha87,Pan03} are also concerned with the basis of
the theory). In \cite{Sch07}, the variational basis of TDDFT proposed by Runge and Gross \cite{Run84} was not only challenged but seemingly refuted. To be more specific, Schirmer and Dreuw claimed that the variational derivation of the time-dependent SKS equations in   \cite{Run84} is not valid due to an ill-defined action functional proposed there. A non-variational treatment would also encounter difficulties, since in this case the SKS system would permit one to reproduce, but not predict, the exact electron density. 

Two contributions involving the present authors \cite{Nie08,Hol08} have been motivated by the criticism in \cite{Sch07} of the RG work. Both of the contributions accept the challenges of the RG proof, but do not require one to abandon the RG conclusion nonetheless. Let us start by summarizing the content of \cite{Nie08}, because this is very specifically focused on time-dependent theory, whereas \cite{Hol08}, though also motivated by the challenges in \cite{Sch07}, is basically dealing with time-independent DFT. 

\subsection{Solvable example of a family of two-electron model atoms
  with general inter-fermion interaction: dynamical generalization}
\label{moshi}
As brief background to the above example, Holas, Howard and March
(HHM) \cite{Hol03}, obtained analytical solutions for ground-state
properties of a whole family of two-electron spin-compensated
harmonically confined members characterized by a given interfermionic
potential energy $u(r_{12})$ (See also \cite{Sam90,Sam91}). In
\cite{Nie08} a start is made on the dynamic generalization of the
harmonic external potential. In the above context, a simplified
expression is obtained for the time-dependent electron density for
arbitrary inter-particle correlation, which is completely determined
by a one-dimensional non-interacting Hamiltonian. That such a
construction is generally possible has been shown by Qian and Sahni
\cite{Qia98,Qia00}, but follows also from the harmonic potential
theorem \cite{Dob94,Sah04} for this specific example. Furthermore for the simplest case: the Moshinsky atom  \cite{Moc68}, where the interaction $u(r_{12})$ is also harmonic, a closed solution for the Fourier transform of the density, namely the time-dependent atomic scattering factor, is found. 

To summarize the essence of the time-dependent density  $n(\rb,t)$ calculation in \cite{Nie08}, from the above model, we take the special but nevertheless important case of a system which is in its ground-state at $t=0$. After generalizing the static separation of center of mass (CM) and relative motion (RM) to the dynamic example under consideration, the above assumption at  $t=0$ leads to the square of the CM wave function as the simple Gaussian form
\begin{equation}
 \label{sm}
  |\psi_{000}^{\text{CM,3D}}(\co,t)|^2 = \frac{1}{a^3_{\text{CM}}(t) \pi^{3/2}} \exp\left(-\frac{c^2}{a^2_{\text{CM}}(t)}\right),
\end{equation}
where the time dependence is determined by the length scale  $a_{\text{CM}}(t)$ of the oscillator.

For the Moshinsky example \cite{Moc68} the time dependent atomic scattering factor $f({\bf k},t)$, defined by
\begin{equation}
\label{scat}
  f({\bf k},t) = \int n({\bf r},t) e^{i{\bf k} {\bf r}} d{\bf r},
\end{equation}
is the convenient tool. The total scattering factor turns out, for $u(r_{12})= -\frac{1}{2} {\cal K} r_{12}^2$, to have the form
 \begin{equation}
  \label{moc}
 f_{\text{tot}}^{\cal K}(k,t) = 2
 f_{\text{\text{CM}}}(k,m_{\text{cm}}\dot{\phi}(t))\, f_{\text{\text{CM}}}(k/2,\tilde{m}_{\text{cm}}\dot{\tilde{\phi}}(t)).
\end{equation}
where $a_{\text{CM}}(t)$  entering eqn(\ref{sm}) is related to $m_{\text{cm}}\dot{\phi}(t)$ in eqn(\ref{moc}) by 
  \begin{equation}
    \label{acm}
    a_{\text{CM}}(t) = \frac{1}{m_{\text{cm}} \dot{\phi}(t)}.
  \end{equation}
Or more generally, the dynamic generalization of the static HHM density is obtained in \cite{Nie08} in terms of the time-dependent relative motion wave function as 
 \begin{eqnarray}
  \label{dnsf}\nonumber
  n({\bf r},t) &=& \frac{8}{\sqrt{\pi}}
  \exp(-\frac{r^2}{a^2_{\text{CM}}(t)}) \\\nonumber &&\times
\int_0^\infty dy\,\, y^2  \exp(-\frac{y^2}{4}) 
 \left|\psi_{000}^{\text{RM,3D}}(a_{\text{CM}}(t) y,t)\right|^2 
 \\ && \times \frac{\sinh(r
   y/a_{\text{CM}}(t))}{(r y/a_{\text{CM}}(t))}.
\end{eqnarray}

Though this is an admittedly simplistic two-electron correlated
time-dependent problem, the time-dependent density $ n({\bf r},t) $
can be got via a one-body time-dependent potential $V({\bf r},t) $,
thereby supporting the original RG assertion. There is no conflict
either, we hasten to add, with the Schirmer-Dreuw conclusions. These
authors, in spite of questioning the RG derivation \cite{Run84} of the
SKS equations, nowhere claim to have disproved this important
assertion! On the contrary, an alternative proof is provided in
\cite{vanL} that the mapping is indeed valid.

To conclude this sub-section, we stress that a correlated two-electron
example proposed in the static limit in \cite{Hol03} has been solved
exactly in the dynamic generalization in which the system is in its
ground state at time $t=0$. In particular, eqn(\ref{dnsf}) allows
the time-dependent density $ n({\bf r},t)$ of the correlated dynamical
problem to be reduced to the single-particle problem of calculating,
probably numerically, from a one-body time-dependent Schr\"{o}dinger
equation, the relative motion wave function
$\psi_{000}^{\text{RM,3D}}$.
 This wave function, though calculated from a one-body equation, involves the sum of the
harmonic confinement potential and the interparticle potential energy
$u$. Having obtained the exact density, also the generating SKS
potential may be obtained by a straightforward, but numerically non-trivial, inversion of the SKS
equations in this two-electron case \cite{lei,ami}. The one-to-one
correspondence of densities and potentials in the non-interacting case
can hence be verified quite explicitly for this specific example.

\subsection{Linear response theory and its inversion}
\label{causal}
Since the comments \cite{Hol08} are basically time-independent, we shall defer these to follow a brief discussion of the important linear response function in time-dependent theory, already to the forefront in the discussion in \cite{Mar72} and \cite{Mar73}. For the non-interacting SKS system generated by potential $V({\bf r},t) $ with first order self-consistent change $\Delta V({\bf r},t) $, with corresponding response function $\chi_s(\rb,t,\rp,t')$, we have for the density change $\Delta n({\bf r},t) $ the formal result
\begin{equation}
\label{res}
\Delta n({\bf r},t) = \int dt' \int \dr' \chi_s(\rb,t,\rp,t') \Delta V(\rp,t').
\end{equation} 
Here the first order change $\Delta V$ in the (now time-dependent) SKS potential is given by 
\begin{equation}
\label{fxc}
\Delta V(\rb,t) = V_\text{ext}(\rb,t) + \int \dr' \frac{\Delta n(\rp,t)}{|\rb-\rp|} +  \int dt' \int \dr' f_{xc}(\rb,t,\rp,t') \Delta n(\rp,t'), 
\end{equation}
where $f_{xc}$ is the as yet unknown, exchange-correlation kernel
\cite{gro95,gro96}. 

In this last cited references by Gross and coworkers, the issue of causality was first raised,
which was also taken up later by Amusia and Shaginyan \cite{Amu98,Amu01} and
Harbola \cite{Har99,Har01,Har02}. The question posed is quite general:
namely whether the potential depends on the density in a causal
manner. This, we believe, is another question important to the
foundations of TDDFT under discussion here. 

The causality issue is important in its own right, but has also
implications for the variational formulation of TDDFT: The
action principle proposed by Runge and Gross \cite{Run84} leads to a
symmetric and hence unexpected non-causal form of the inverse of the
response function $\chi$, giving rise to what has been termed the {\em
  symmetry-causality paradox} (see also \cite{Coh05}). Different reformulations of the action
principle have appeared in the literature, e.g., \cite{Raj96,Lee98}, up to a
recent contribution by Vignale in which the principle of least action
in its conventional form 
is abandoned \cite{Vig08}.

Returning to the question of causality, we follow the treatment of Amusia and Shaginyan \cite{Amu01} who write the external potential in terms of the density, by invoking the many-body linear response function $\chi$ (in contrast to the one-body $\chi_s$ used in eqn(\ref{res}) above). With  the external potential $V_\text{ext}(\rb,t)$ one can write 
\begin{equation}
\label{har}
\Delta n(\rb,t) = \int dt' \int \dr' \chi(\rb,t,\rp,t') V_\text{ext}(\rp,t'). 
\end{equation}
In \cite{Amu01}, it is assumed that eqn(\ref{har}) can be changed to the Volterra integral equation (see also \cite{Har02}):
\begin{equation}
\label{vol}
\frac{1}{K(t)} \frac{\partial^2 \Delta n(t)}{\partial t^2}  = V_\text{ext}(t) + \int dt' \frac{1}{K(t)} \frac{\partial^2 \chi(t,t')}{\partial t^2}  V_\text{ext}(t'), 
\end{equation}
where the spatial variables are omitted for clarity of notation. The function $K(t)$ entering eqn(\ref{vol}) is defined by \cite{Amu01}
\begin{equation}
\label{kde}
K(t) = \left.  \frac{\partial \chi(t,t')}{\partial t'} \right|_{t=t'}.
\end{equation}
The solution of eqn(\ref{vol}) has the form \cite{Amu01,Har02}
\begin{equation}
\label{kex}
 V_\text{ext}(t) = \frac{1}{K(t)} \frac{\partial^2 \Delta n(t)}{\partial t^2} + \int dt' R(t,t') \frac{1}{K(t')}  \frac{\partial^2 \Delta n(t')}{\partial t'^2}, 
\end{equation}
where $R(t,t')$ vanishes for $t < t'$. Harbola \cite{Har02} argues
from the above treatment that while the density depends on the
potential in a causal manner, the reverse is not true, leaving
no room for a symmetry-causality paradox. On the other side, Amusia and Shaginyan
\cite{Amu98,Amu01} as well as Cohen and Wasserman \cite{Coh05} provide
arguments that causal inverse response functions indeed exist. Our present contention is
that all-important for a decisive answer to the causality
question resides in the mathematical properties of
eqn(\ref{kex}). This is an important future area therefore for rigorous
mathematical physics.

\subsection{Two further comments pertaining to the Schirmer-Dreuw study}
At this point, we return to the comments of Holas et al.~\cite{Hol08}
on the Schirmer-Dreuw study \cite{Sch07}. As already mentioned above,
in \cite{Hol08} the two particular subjects tackled are concerned with
the original, time-independent DFT but have implications also for the
time-dependent generalization. In the present section, we merely summarize the relevant points in  \cite{Hol08} for the present context.

The first focus of  \cite{Hol08} was to answer in a positive fashion a
question posed in \cite{Sch07}. Whether a local operator can be
reconstructed from knowledge of its particle-hole (p-h) matrix
elements when the number of particle states exceeds one is the essence
of the question. It arises in the context of a linear response
treatment within TDDFT, in which only p-h matrix elements of the
perturbing operator appear in the relevant equations, although an exact
many-body approach requires also the knowledge of p-p and h-h
elements. It turns out that there is no conflict with the exactness of
TDDFT due to this apparent loss of information \cite{Sch07}.  
Schirmer and Dreuw formulate and prove the theorem that a local (multiplicative) operator $u=u(\rb)$ is uniquely determined to within a constant by its p-h and h-p matrix elements with respect to a complete one-particle basis and {\em any} partitioning of that basis into occupied and unoccupied one-particle orbitals. But in answer to the question to whether it is possible to reconstruct a local operator if only its p-h matrix elements are given, in \cite{Sch07} it is remarked that it seems not possible except for a special case when the number of occupied particle states ($n$) equals one. It is stressed by Holas  et al.~\cite{Hol08} that there is a positive answer for any $n$ given in Sec. III of the work of Holas and Cinal \cite{Hol03}.

As a second focus arising from the work in \cite{Sch07}, Holas et
al.~\cite{Hol08} refer to the differential equation satisfied by the
density amplitude $n(\rb)^{1/2}$, involving the concept of the Pauli
potential. While conceptually satisfying, in \cite{Hol08} some reasons
are set out why the implementation of such a {\em radical Kohn-Sham
  scheme} \cite{Sch07} is
presently hardly computationally competitive with that based on
Slater-Kohn-Sham (SKS) orbitals. Being numerically demanding already
for the static case, we expect the computational scheme discussed by
Schirmer and Dreuw to be even more involved in the full time-dependent
case.

\section{The differential virial theorem in time-dependent theory}
We turn next  to an important result for DFT: namely the differential virial theorem (DVT). Following the earlier study of March and Young \cite{Mar59} on the idempotent Dirac density matrix via its equation of motion, which led then in one-dimension to a result which they termed the differential form of the virial theorem, Holas and March \cite{Hol95} some four decades later established the DVT in three dimensions with full account also taken of electron-electron interactions. Ref.~\cite{Hol95} has a fairly direct generalization to time-dependent theory \cite{Qia98,Qia00,Qia01,Har02} and leads to the result (see also \cite{Mar72}):
 \begin{equation}
\label{dvt}
\frac{\partial^2 n(\rb,t)}{\partial t^2} = -\frac{1}{4} \nabla^4
n(\rb,t) +\nabla {\bf z}(\rb,t) +2\nabla \int
\dr'   n_2(\rb,\rp,t)\nabla u(\rb,\rp) + \nabla \left[
  n(\rb,t) \nabla V_\text{ext}(\rb,t)\right].
\end{equation} 
Here $n_2(\rb,\rp,t)$ is the pair density, $u(\rb,\rp)$ is the electron-electron repulsion potential energy, while $z(\rb,t)$ is a vector field defined from the kinetic energy density tensor $t_{\alpha\beta}(\rb,t)$ following Holas and March \cite{Hol95}. To be a little more specific, eqn(\ref{dvt}) is, in essence a combination of the DVT and the continuity equation relating density $n(\rb,t)$ and current density ${\bf j}(\rb,t)$, namely
 \begin{equation}
\label{con}
\frac{\partial n(\rb,t)}{\partial t} + \nabla  {\bf j}(\rb,t) = 0,
\end{equation} 
the latter being already invoked in the early work of March and Tosi
\cite{Mar72} on TDDFT. What we want to stress is that eqn(\ref{dvt})
can be employed at least in principle to construct the applied
(external) potential from the density $n(\rb,t)$ plus the initial
conditions, a matter already touched on in
Sec.~\ref{intro}.  Prerequisites for such a construction are accurate
approximations for the kinetic energy density tensor and the
correlated pair density. Efforts to obtain the latter in terms of first-order
density matrices \cite{Val92}, should be very useful in this respect.

Rewriting eqn(\ref{dvt}) for the non-interacting SKS system, we have
 \begin{equation}
\label{ksdvt}
\frac{\partial^2 n(\rb,t)}{\partial t^2} = -\frac{1}{4} \nabla^4
n(\rb,t) +\nabla {\bf z}_s(\rb,t)  + \nabla \left[
  n(\rb,t) \nabla V(\rb,t)\right],
\end{equation}
which requires only the knowledge of non-interacting vector field
$z_s(\rb,t)$ \cite{Qia00} to extract the Kohn-Sham potential $V(\rb,t)$ from a
given density $n(\rb,t)$. This should be helpful in cases where direct
inversion of the Kohn-Sham equations is impossible (cf.~sub-section
\ref{moshi}).

In order to obtain expressions for the
exchange-correlation potential as {\em functional} of the density,
admittedly, the pair density $n_2(\rb,\rp,t)$ still enters in the
subtraction of $V_\text{ext}(\rb,t)$ from $V(\rb,t)$. In
Ref.\cite{Hol95} dealing with the time-independent problem, the
correlated pair density and kinetic energy density tensor were replaced by their non-interacting
counterparts, which lead after a combination of the analogues of
eqn(\ref{dvt}) and eqn(\ref{ksdvt}) to an exchange-only approximation
of the exchange-correlation potential beyond the Slater form. A
similar proceeding is expected to succeed also in the present
time-dependent case.  As discussed by Qian and Sahni
\cite{Qia98,Qia00,Qia01} in their derivation of eqn(\ref{dvt}) and (\ref{ksdvt}), the combination
of the above equations is also important from a more formal point of view,
since it allows to disentangle several contributions to the
exchange-correlation functional and interpret these in physical terms.

\section{Shortcomings of present calculations by TDDFT on charge-transfer excitations}
It is highly relevant to the foundations of TDDFT that current usage leads to substantial errors for charge-transfer excited states \cite{Dre03,Toz99,Dre05}. Usually, the excitation energies are severely underestimated. Furthermore, the potential energy curves   of such charge transfer states do not display the known $1/R$ dependence along a charge-separation coordinate $R$ \cite{Dre03b,Dre04,Toz03}. 

A long-term solution to this problem may well lie in the use of time-dependent current density functional theory, which has recently been implemented \cite{Faa03} (see also \cite{Mar72,Gro90,Mar06}). It turns out, as discussed in \cite{Faa03} that a correct description of charge transfer excited states requires non-locality, and current density functionals have this property. However, the high computational cost of such a current density approach raises doubts as to whether this route will be applicable to large molecules in the foreseeable future.

\section{Summary}
Serious criticisms of the foundations of TDDFT have recently been made
\cite{Sch07}, the focus being on parts of the Runge-Gross work
\cite{Run84}. The example set out in
sub-section \ref{moshi}, and culminating in eqn(\ref{dnsf}), shows in
an admittedly simple time-dependent problem with an exact solution,
that the time-dependent density can be correctly calculated from a
one-body time-dependent potential in support of the mapping theorem
\cite{Run84}.  

We stress here again that the mapping theorem is not challenged
in \cite{Sch07}. The major point of the Schirmer-Dreuw study 
is rather the claim that the time-dependent Kohn-Sham approach has no predictive power
due to the lack of a valid variational principle.
However, once the time-dependent density, discussed in some detail in
sub-sections \ref{moshi} and \ref{causal} above, is obtained by some
other means, and the exact time-dependent exchange-correlation
potential-functional would be known, the exact time-evolution of the
electron density of the interacting system can be reproduced by the
time-dependent Kohn-Sham equations. Since this manuscript was
completed, a comment \cite{Mai08} 
and reply \cite{Sch08b} on the  Schirmer-Dreuw study
\cite{Sch07} have been published.

Further matters discussed involve questions of the causality of the
potential in TDDFT, first raised by Gross and
co-workers \cite{gro95,gro96} and subsequently discussed by  Amusia
and Shaginyan \cite{Amu98,Amu01} and Harbola \cite{Har02}, and of the need to face the additional complications of current density theory for a specific class of excitations, namely charge-transfer excited states. This is because of the fundamental need for non-locality, which is correctly embedded in current density theory. Questions then arise as to the feasibility of application of such an approach to large molecules, because of the high cost.

\begin{acknowledgements}
We wish to acknowledge that the present article was brought to fruition during a visit of both authors to the Division of Molecular Biophysics at the German Cancer Research Center (DKFZ). It is a pleasure to thank Professor S. Suhai for generous hospitality and for arranging a Scholarship to support the visit. Finally NHM thanks Prof. A. Rubio for valuable discussions pertaining to charge-transfer excitations.    
\end{acknowledgements}


\end{document}